**Leveraging Social Media Data to Identify Factors Influencing Public Attitude Towards Accessibility, Socioeconomic Disparity and Public Transportation**

**Khondhaker Al Momin**
Ph.D. Student
School of Civil Engineering and Environmental Science
University of Oklahoma
202 W. Boyd St., Norman, OK 73019-1024
Email: momin@ou.edu

**Arif Mohaimin Sadri, Ph.D.**
Assistant Professor
School of Civil Engineering & Environmental Science
University of Oklahoma
202 W. Boyd St., Norman, OK 73019-1024
Email: sadri@ou.edu
(Corresponding Author)

**Md Sami Hasnine, Ph.D.**
Assistant Professor
Department of Civil and Environmental Engineering
Howard University
2400 6th St NW, Washington, DC 20059
Email: mdsami.hasnine@howard.edu

\*\* All authors contributed equally

**ABSTRACT**
This study proposes a novel method to understand the factors affecting individuals' perception of transport accessibility, socioeconomic disparity, and public infrastructure. As opposed to the time consuming and expensive survey-based approach, this method can generate organic large-scale responses from social media and develop statistical models to understand individuals' perceptions of various transportation issues. This study retrieved and analyzed 36,098 tweets from New York City from March 19, 2020, to May 15, 2022. A state-of-the-art natural language processing algorithm is used for text mining and classification. A data fusion technique has been adopted to generate a series of socioeconomic traits that are used as explanatory variables in the model. The model results show that females and individuals of Asian origin tend to discuss transportation accessibility more than their counterparts, with those experiencing high neighborhood traffic also being more vocal. However, disadvantaged individuals, including the unemployed and those living in low-income neighborhoods or in areas with high natural hazard risks, tend to communicate less about such issues. As for socioeconomic disparity, individuals of Asian origin and those experiencing various types of air pollution are more likely to discuss these topics on Twitter, often with a negative sentiment. However, unemployed, or disadvantaged individuals, as well as those living in areas with high natural hazard risks or expected losses, are less inclined to tweet about this subject. Lack of internet accessibility could be a reason why many disadvantaged individuals do not tweet about transport accessibility and subsidized internet could be a possible solution.

**BACKGROUND AND MOTIVATION**
The transportation system is vital in providing travel options and impacting individuals' quality of life (*1*). With the increasing racial and ethnic diversity due to a rising influx of immigrants, there is a need for alternative approaches in transportation planning to address equity issues in transportation experienced by travelers from different backgrounds (*2*). Transportation accessibility pertains to the ease of reaching societal and economic opportunities like education, employment, and healthcare, which is often challenging for those relying on transit or active transportation (*3*). While diversity in transportation refers to developing infrastructure and services to accommodate varying backgrounds, income levels, and abilities (*4*); equity ensures fair access to infrastructure and services for everyone, including the equal access to public transport across all neighborhoods and affordable options (*5-9*). Inclusion seeks to create an environment where everyone has a sense of belonging that can be achieved by designing infrastructure and services that catered to individuals with disabilities, such as wheelchair ramps, lifts, audio-visual announcements, and accessible parking (*10,11*).

Social equity in transportation has recently attracted attention from researchers, transportation agencies, and the private sector, aiming to reduce adverse impacts on underrepresented and marginalized travelers and workers (*1,12-22*). This focus on social equity in transportation arose from efforts to address systemic underinvestment and underrepresentation in all aspects of infrastructure life-cycle phases, including planning, design, construction, operations and maintenance. This further escalated due to increased engagement and participation from underrepresented groups and marginalized travelers from the bottom-up. This is evident from the Transportation Research Board (TRB) that recently identified complementary pathways to achieve equity goals and set priorities at the local, tribal, state, regional, and federal levels (*23*). However, earlier focus on diversity (*24,25*) has now shifted to equity considerations to reflect more on inequities due to ethnicity, language and income among others (*1-7,10,11,15,26-31*). For instance, vulnerable communities often face financial limitations preventing them from living in areas with better infrastructure or owning a vehicle. These constraints, coupled with physical limitations or choices not to drive, limit their access to social and economic opportunities. This often results in poor living conditions, economic stagnation, high unemployment rates, social isolation, and enduring social inequalities (*2,3,15*).

While survey-based approaches are conventionally used to assess social equity and accessibility in transportation (*9,27*), recent studies have adopted alternative techniques. For example, one study proposed a two-fold approach to design an equity audit tool based on transportation needs in a community (*32*). However, to complement survey-based approaches to collect behaviorally enriched data on social equity and accessibility in transportation, this study presents social media data as a viable alternative with excellent statistical power. In particular, geo-tagged social media data provides a great opportunity to integrate these datasets with nationally representative household surveys such as National Household Travel Survey (NHTS), American Community Survey (ACS) and national databases such as Social Security Administration (SSA) and census data. Such integration allows testing different household and community-level factors in addition to the ones obtained through the online interactions of public in social media systems. For example, transportation researchers have extensively used social media data (e.g., Twitter) in various applications (*33-47*). However, the literature is inconclusive how to systematically extract and process social media data to capture public concerns and needs regarding their transportation systems in the neighborhood and reveal the factors that influence public attitude towards contemporary issues.

This study utilized large-scale social media data from Twitter to analyze individuals' perception of transport accessibility, socioeconomic disparity, and public infrastructure in New York City (NYC). The study aims to reveal the factors influencing the likelihood of individuals tweeting about transport accessibility, socioeconomic disparity, and public transport infrastructure. This study introduces a novel methodology that enhances the transportation literature by overcoming limitations of social media data identifying socio-demographic information of social media users such as race and gender, followed by integrating the geotagged tweets with other databases such as SSA, Census, ACS among others. Finally, a series of econometric models have been estimated to understand the overall quality of the generated dataset.

**DATA DESCRIPTION**
The study chose New York City for studying transport accessibility, socioeconomic disparity, and public infrastructure due to its unique characteristics and comprehensive data availability. As one of the most densely populated, and diverse (Figure 1)cities in the United States, New York City boasts a robust public transportation system, including the largest subway system in terms of number of stations and an extensive network of bus routes (*48*). This makes it an excellent location for studying transport accessibility and the impact of public transportation on urban life. The city's socioeconomic disparity is also stark, with some of the wealthiest individuals and neighborhoods in the country co-existing alongside areas of significant poverty (*49*). This provides a rich context for investigating the relationship between socioeconomic status and access to resources and opportunities. In terms of public infrastructure, New York City's vast array of public facilities, ranging from parks and schools to hospitals and libraries, offers a broad scope for studying the distribution and accessibility of public infrastructure (*50*). Furthermore, the city's government agencies regularly publish extensive data on these topics, facilitating in-depth research and analysis.

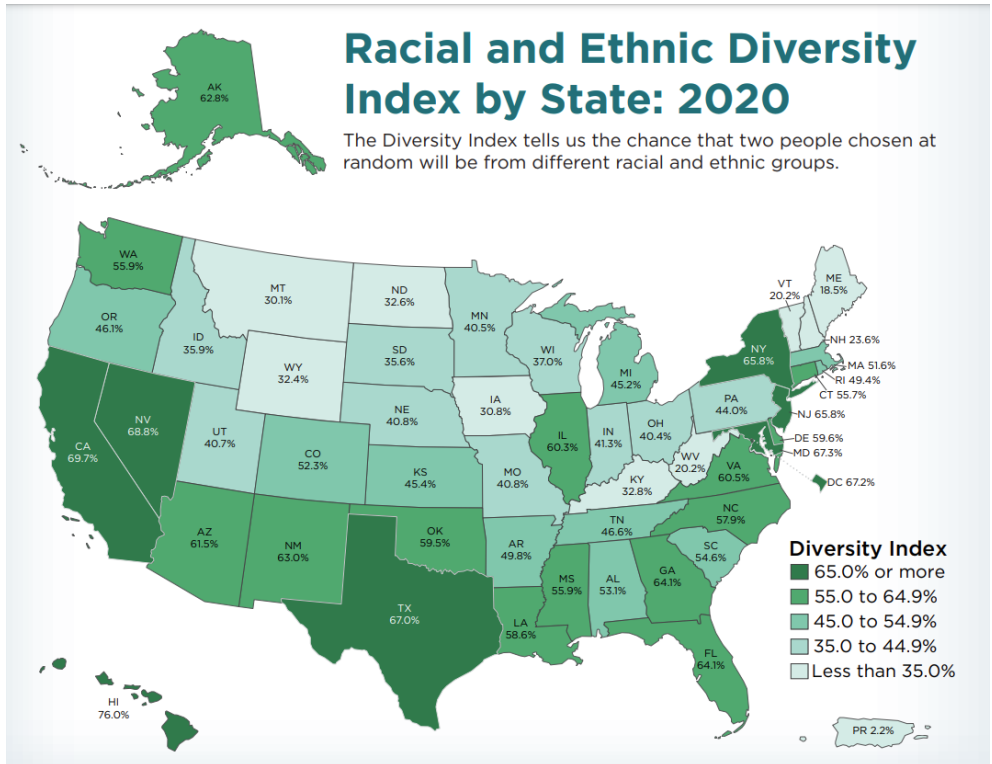

*Figure 1: Racial and Ethnic Diversity Index by State (source https://www.census.gov/library/visualizations/2021/dec/racial-and-ethnic-diversity-index.html)*

*Twitter Data*

In this study, the data collection process relied on the Twitter Academic Application Programming Interface (API) (*51*). Academic API has access to the entire historical tweets based on specific search queries through its full-archive search endpoint (*52*). To facilitate the data collection process, the study used the Python programming language along with relevant Python libraries. For the purpose of the study, a geolocation-based search inquiry was strategically employed to target only geotagged tweets within the designated study area, as depicted in **Figure 2**. This allowed the study to extract all geotagged tweets between the time frame of March 15, 2020, to May 15, 2020, resulting in a dataset comprising a total of 1.18 million tweets, originating from approximately 41,580 unique users. Tweet data retrieved via the academic track API also provides ancillary data elements including user identification number, username, profile description, and geographical coordinates associated with the originating tweet. The spatial distribution of the collected tweets is depicted in the figure below.

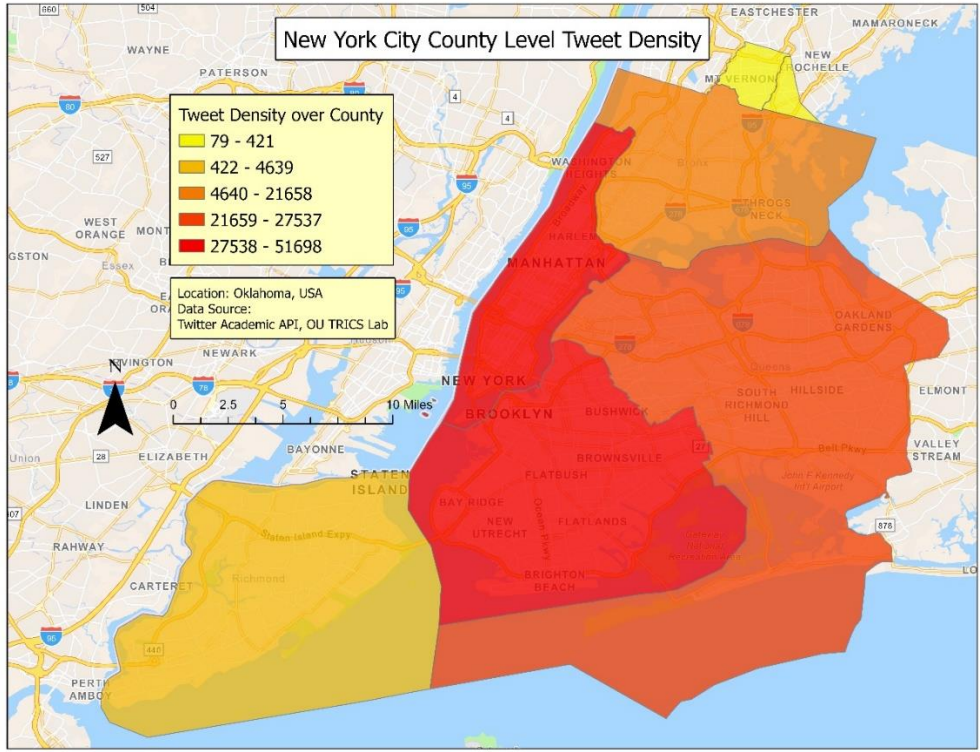

**Figure 2: Study area**

This study took into account tweets related to transportation accessibility, socioeconomic disparity, and public infrastructures. The relevance of a tweet was ascertained based on the identification of specific tokens within the tweet. An in-depth exploration of the relevance filtering process, including its methodologies and significance, can be found in reference (*53*).

*Social Security Administration Data*
Social Security Administration (SSA) (*54*) has been diligently gathering all names submitted through social security card applications for births in the United States since 1879. In order to analyze the gender demographics of Twitter data, the study utilized the first names collected from the SSA website. As of 2021 the SSA data encompassed 63,152 unique male names and 37,212 unique female names, offering a substantial and reliable foundation to train machine learning models to predict gender from the first names.

*Census Data*
The United States Census Bureau compiles annual estimates of race and Hispanic origin proportions for every county in the nation, based on respondents' surnames. These estimates stem from the most recent decennial census data and are further adjusted to account for population changes like deaths, births, and migration since that time. To streamline this study, the complex topic of race/ethnicity has been categorizing it into four groups: Asian, Black, Hispanic, and White. The study obtained last names associated with different racial and ethnic backgrounds from the census bureau (*55*) in order to ascertain a user's race/ethnicity from the Twitter users' name. This comprehensive dataset comprises a total of 162,254 surnames representing various races and ethnicities. By employing this data, this study aims to gain insights into the racial composition of Twitter users, acknowledging the importance of precise categorization in the research.

*Societal and Economic Drivers*

Socioeconomic characteristics shows a high degree of variation across different counties and block groups. NYC has five counties (New York, Kings, Queens, Bronx, Richmond) and around 6807 census block groups. To comprehend the effect of these variations on public sentiments towards various transportation-related options, the authors of this study collected data about five pertinent socioeconomic factors at the census block group level from the American Community Survey (ACS)*(13)* database. The collected factors at the census block group level are:

- *Per Capita Income (PCI):* Average income per person. A higher PCI might mean more discretionary income for travel.
- *Mean Travel Time to Work*: Average commute time. Longer times may highlight inefficiencies in public transport or distance from work areas.
- *Proportion of People Living Under the Poverty Limit*: Percentage of population below the poverty threshold. Higher rates may imply reliance on cheaper or public transportation.
- *Unemployment Rate*: Percentage of jobless people actively seeking work. High unemployment may impact travel behaviors, such as less commuting.
- *High School Completion Rate*: Percentage of people with at least a high school education. Higher education levels can correlate with better job stability and more travel flexibility.
- *Hazard Exposure Variables*: Hazard exposure variables e.g., PM2.5, Diesel Particulate Matter, Proximity to Hazardous Sites measure environmental risks including air pollutants and proximity to hazardous sites.
- *Health Indicator Variables:* Indicate the prevalence of specific health conditions like asthma, diabetes, and heart disease among adults. Higher prevalence rates suggest an increased burden of disease, which can be influenced by a combination of socioeconomic, environmental, and healthcare access factors.
- *Housing-Related Variables:* Housing related variables such as Housing Burden, Lead Paint Indicator, Median Value of Housing Units provide insights into living conditions, including housing affordability, potential lead exposure, and the percentage of income spent on housing.
- *Social Vulnerability Variables:* Social vulnerability e.g., Linguistic Isolation, Low Life Expectancy, Individuals Below Federal Poverty Line highlight social challenges like linguistic isolation, low life expectancy, and the percentage of individuals living below the federal poverty line.

**Data Preparation**

The study involved several steps in data preparation. Firstly, Twitter data underwent preprocessing, during which tweet texts, user information, profile description and location data have been extracted and cleaned. This involved removing noise such as Hypertext Markup Language (HTML) tags, emoticons, and stop words, and tokenizing the data into smaller units. The study also integrated demographic and socioeconomic factors with the Twitter data to create a comprehensive dataset. The geolocation (latitude and longitude) of tweets were determined by reverse geocoding using the Census Geocode python package. Lastly, the dataset was refined to only include tweets related to transportation accessibility only by using relevant filtering techniques as described by Kryvasheyeu et al. (*56)*. Following the relevance filtering, 36,098 tweets discussing about transportation accessibility were extracted.

## METHODOLOGY
### Race and Gender Identification
The objective of the gender-race (GR) model is to predict gender from first names and race from last names. The development of the GR model is comprised of several steps. The first step is to collect first names SSN data (n=100,364) and surnames from the United States Census Bureau (n=162,254). A comma separated value (CSV) file containing last names (for race prediction) and first names (for gender prediction), along with their respective race/gender labels, was used as the primary source of data.

After collecting the data, it was preprocessed to ensure its suitability for further analysis. The last names were extracted from the CSV file and loaded into a data frame. Next, a function was created to calculate the occurrences of each letter in a name. These letter counts were then utilized to create an alphabet matrix accompanied by additional race and gender labels for each name. The next stage involved feature engineering to create a model-ready dataset. The alphabet matrix was created by transforming the letter counts into concatenated strings associated with each name. Each string was then linked with a race label from four categories: Asian, Black, Hispanic, and White. Similarly, each string was associated with a gender label from two categories: Male and Female.

*Model Training and Evaluation:*
The alphabet matrix was segregated into a feature matrix X, containing the letter counts, and a target vector y, containing the race and gender labels. The dataset was then divided into training and testing sets, adhering to a 70-30 train-test split. This split provided sufficient data for model training while ensuring a robust evaluation. Various machine learning classifiers such as Random Forest (RF), Decision Tree (DT), K-Nearest Neighbor (KNN), Support Vector Machine (SVM), and Naïve Bayes (NB) were used to train the models on the training data to predict race and gender labels from letter counts. The model's accuracy was evaluated on the test set, and the out-of-bag accuracy was computed to mitigate the risk of overfitting. Additionally, ten-fold cross-validation was performed to estimate the skill of the machine learning models on new data and to provide robustness against the randomness of the training and testing data split.

The performance of the models have been evaluated using various metrics. The accuracy score provided an indication of how effectively the model predicts race labels based on last names. To visualize the number of correct and incorrect predictions for each race category, a confusion matrix was generated. Additionally, a classification report was generated, which included precision, recall, F1-score, and support for each race category. These metrics provided a comprehensive understanding of the model's performance across different categories and helped identify areas for improvement.

### Bidirectional Encoder Representations from Transformers (BERT) Model
Bidirectional Encoder Representations from Transformers (BERT) is a revolutionary pre-trained model that has significantly improved the state of the art for many natural language processing (NLP) tasks (*57*). The model starts its lifecycle in a pre-training phase, using large text corpora like Wikipedia, wherein it learns language structure and semantics via two unsupervised learning tasks: Masked Language Model (MLM) and Next Sentence Prediction (NSP) (*58*). After pre-training, BERT can be fine-tuned for task-specific objectives such as text classification (*59*). The fine-tuning process adjusts the weights in the model to better suit the particular task, leveraging a task-specific dataset. When preparing input for BERT, sentences are tokenized into *subwords* using the BERT *WordPiece* tokenization scheme and then converted into vectors, each being assigned a unique integer or token ID (*60*). The abbreviation *CLS* in the context of BERT stands for *Classification*. The [*CLS*] token is a special symbol added to the beginning of an input sentence (or sequence of sentences) during preprocessing, and its final hidden state is used as the aggregate sequence representation for classification tasks.

The model also utilizes Positional and Segment Embeddings to represent the position of each word in the sentence and distinguish different sentences, respectively (*61*). The input sequence is fed into BERT's core, which consists of a multi-layer bidirectional Transformer encoder (*62*). Unlike traditional models, BERT processes all tokens concurrently, employing self-attention mechanisms to comprehend the influence of each token on others, irrespective of their position in the sentence (*63*). Given a sequence of token embeddings $\mathbf{Y} = \{y_1, y_2, ..., y_n\}$ in a particular layer, the self-attention mechanism first computes a set of Query (Q), Key (K), and Value (V) vectors for each token $x_i$ through learned linear transformations:

$$Q_i = y_i\, W_{xx} \quad \text{(Query transformation)}$$
$$K_i = y_i\, W_{xk} \quad \text{(Key transformation)}$$
$$V_i = y_i\, W_{xv} \quad \text{(Value transformation)}$$

Where $W_{xx}$, $W_{xk}$, and $W_{xv}$ are the learned weight matrices. The attention score for a pair of tokens i and j is computed as the dot product of their query and key vectors, normalized by the square root of the dimension of the key vectors ($d_k$):

$$\text{Attention}(Q_i, K_j) = \text{SoftMax}\left(\left(\frac{Q_i . K_j}{\sqrt{d_k}}\right)\right)$$

This score indicates how much focus to place on the $j^{th}$ token while encoding information for the i-th token. Then, the output vector for each token is computed as a weighted sum of all value vectors, where the weights are the attention scores:

$$\text{Output}_i = \Sigma_j\, (\text{Attention}(Q_i, K_j) * V_j)$$

In practice, these computations are all done in parallel using matrix operations, and the self-attention mechanism actually computes multiple sets of Q, K, V vectors (called "heads") for each token, concatenating their output vectors together.

These calculations are part of the Transformer architecture on which BERT is based on. In the context of BERT specifically, these computations are performed in a bidirectional manner, allowing each token to attend to all other tokens in the sequence. This is a key factor that enables BERT's strong performance on many NLP tasks. In text classification tasks, the output vector corresponding to the token from the final transformer layer serves as the sequence representation and is subsequently fed into a task-specific classifier. Both the classifier and BERT model weights are adjusted during fine-tuning, training the classifier to make predictions based on the BERT model output. Once the fine-tuning process is complete, BERT can classify new text examples. This involves processing the input text as done during training and then feeding the output of the token into the classifier to generate a prediction. BERT's bidirectional nature, which enables it to understand a word's context concerning all other words in the sentence, marks a distinct advance over previous models.

**Binary Logit Model**
The binary logit model is a type of logistic regression model that is used when the dependent variable (the outcome of interest) is binary, meaning it has two possible outcomes (*64*). The binary logit model can be represented mathematically as follows (*65*):

$$\log\left(\frac{P(Y=1)}{1-P(Y=1)}\right) = \beta_0 + \beta_1 X_1 + \beta_2 X_2 + ... + \beta_n X_n \quad (1)$$

Here, P(Y=1) represents the probability of the event of interest (e.g., a tweet is related to transport accessibility), and $\frac{P(Y=1)}{1-P(Y=1)}$ is the odds of the event happening. The logarithm of the odds (log-odds) is modeled as a linear combination of predictor variables $X_1, X_2, ..., X_n$, with coefficients $\beta_0, \beta_1, \beta_2, ..., \beta_n$. The β values are parameters that the model will estimate, with $\beta_0$ being the intercept term (the log-odds when all predictors are zero), and $\beta_1, \beta_2, ..., \beta_n$ being the coefficients that explain the relationship between the predictors and the log-odds of the event (*66*). Once the logit model is estimated, the probability of the outcome can be obtained using the logistic function:

$$P(Y=1) = \frac{e^{\beta_0 + \beta_1 X_1 + \beta_2 X_2 + ... + \beta_n X_n}}{1+e^{\beta_0 + \beta_1 X_1 + \beta_2 X_2 + ... + \beta_n X_n}} \quad (2)$$

In the context of this study, the binary logit model is used to estimate the likelihood of individuals tweeting about transport accessibility, socioeconomic disparity, and public transport infrastructure issues. The model's goodness-of-fit is evaluated using the adjusted rho-squared against the null model, where the null model is a model with no predictors (just an intercept) (*67*). The resulting probabilities can then be used to make predictions or understand the relationships between the predictor variables and the outcome.

**RESULTS**
**Gender-Race model**
Using gender data obtained from SSN data (n=100,364) and race data from the United States Census Bureau (n=162,254), this study meticulously trained a diverse set of machine learning classifiers such as Random Forest (RF), Decision Tree (DT), K-Nearest Neighbor (KNN), Support Vector Machine (SVM), and Naïve Bayes (NB) models. The following Figure *2* depicts the performance of these machine learning models in their pursuit of accurately predicting gender. Based on the performance metrics, the Random Forest (RF) model appears to be the most suitable for predicting gender (male or female). It achieves the highest average accuracy of 0.84, along with robust performance in precision, recall, and F1-score, indicating a balanced performance in terms of both false positives and false negatives. This suggests that the Random Forest model is the most effective at correctly classifying instances and maintaining a balance between precision and recall, making it the recommended choice for predicting gender.

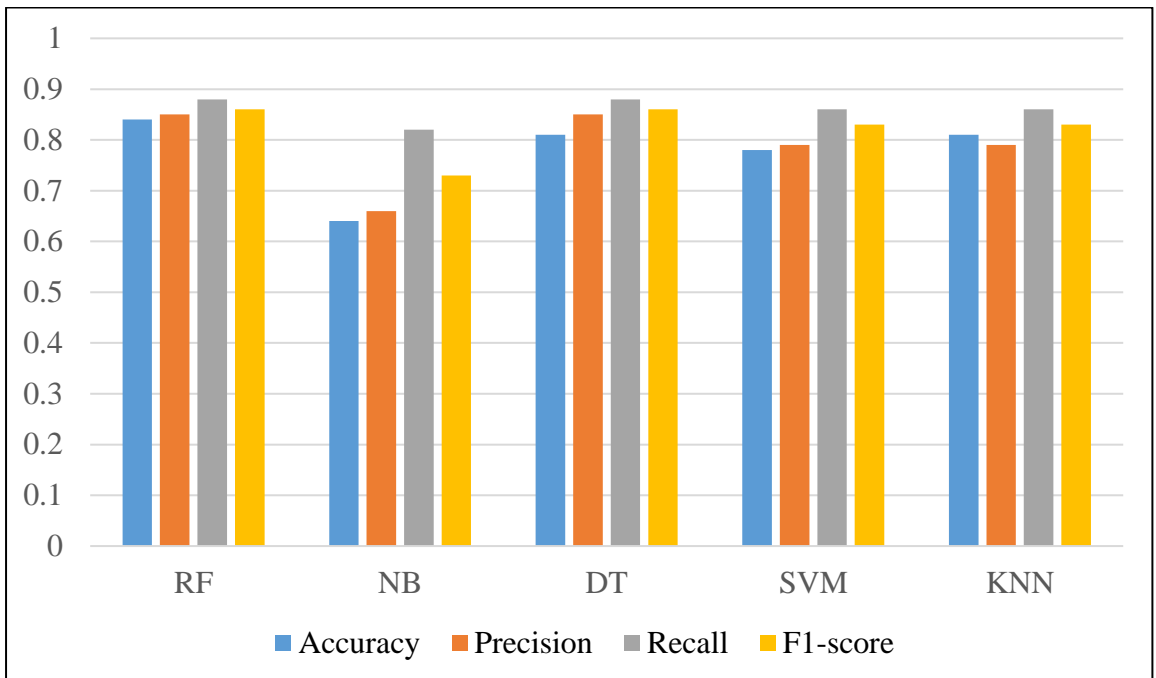

Figure 2: Gender Prediction Model performance values (accuracy, precision, recall, F1 score)

The study found Support Vector Machine (SVM) model is the most suitable for predicting race from the four categories: Asian, Black, Hispanic, and White based on the performance metrics (reference Figure 3). The SVM model achieves the highest average accuracy of 0.84 and also demonstrates superior precision, recall, and F1-score, indicating a strong balance between the number of correct predictions and the ratio of true positive predictions. This balance is crucial when predicting multiple classes, as it ensures the model performs well across all categories. Therefore, considering these metrics and the multiclass nature of the problem, the SVM model has been found suitable for predicting race.

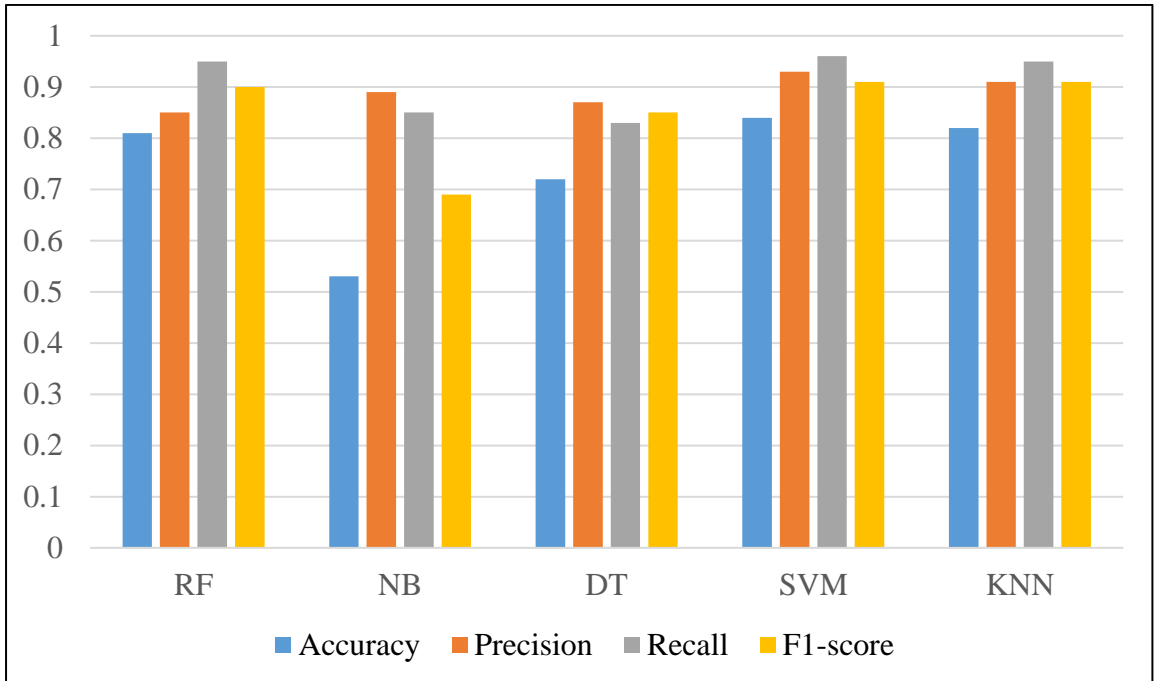

Figure 3: Race Prediction Model performance values (accuracy, precision, recall, F1 score)

**Text Classification**

The study retrieved 36,098 tweets, all relevant to transportation accessibility. This study used topic modeling to determine the optimal number of topics discussed in these tweets. Topic model allows for effective segmentation, understanding, and summarization of such large datasets. With numerous topic modeling approaches, the study chose Latent Dirichlet Allocation (LDA) model (*68*) for the analysis. The rationale behind choosing LDA was to uncover hidden and undiscovered topics related to transportation accessibility, public transportation infrastructure, Socioeconomic Disparity, and others. LDA operates as a generative probabilistic model, in contrast to supervised models that can only identify the known topics they were trained on. It assumes each document has a variable distribution of topics, allowing it to discover novel associations and patterns within the data. This study found six optimum topics with a coherence score of 0.5926. The following ***Table 3*** shows the topics contributing most to the topic and their associated probability in the document. In pursuit of comprehensive analysis, the study categorized the entire dataset into four distinct categories. These categories included *Accessibility*, *Public Transport Infrastructure*, *Socioeconomic Disparity*, and others. ***Table 2*** depicts the sample tweets from *Accessibility, Public Transport Infrastructure,* and *Socioeconomic Disparity*.

To train and test the BERT model, the authors manually labeled 600 tweets from each of these categories, culminating in a total of 2400 annotated tweets. Armed with this precisely labeled dataset, the study proceeded to train the BERT model, leveraging its advanced capabilities in natural language processing. BERT's cutting-edge architecture allowed it to effectively learn from the provided data, grasping intricate patterns and semantic nuances, thus enhancing its understanding of the diverse topics encompassed within the categorized tweets. This robust training process equipped the BERT model to deliver insightful and contextually accurate analyses across the broad spectrum of the dataset's themes.

**Table 1: Topic Model Result**

| Topic Name | Words (probability) in coherent topic |
|---|---|
| Public Transport Infrastructure | station (0.131), line (0.112), infrastructure (0.062), street (0.021), bus (0.018), bothdir (0.010), road (0.009), train (0.009), commuter (0.009) |
| Socioeconomic Disparity | income (0.014), population (0.011), inequality (0.010), racism (0.010), service (0.009), help (0.008), afford (0.008), need (0.007), rich (0.067), poor (0.003), inequity (0.043), equity (0.026), class (0.035), deprived (0.007), social (0.004) |
| Accessibility | accessible (0.020), accommodation (0.016), access (0.016), shelter (0.014), able (0.013), opportunity (0.012), town (0.013), wheel (0.028), seat (0.017), access (0.011), chamber (0.024), free (0.029), park (0.012) |

F1 score, precision, and recall are the common metrics used to assess the performance of a machine learning model, particularly in classification tasks (69). The BERT model, which was meticulously trained using the extensive dataset from this study, has exhibited remarkable performance, boasting an impressive F1 Score of 0.9868. Moreover, the model achieved a Precision of 0.9867, ensuring the accuracy of its positive predictions. Additionally, it demonstrated an outstanding Recall of 0.9869, effectively capturing a vast majority of the relevant instances. These stellar results highlight the efficacy and reliability of the BERT model in handling the tasks presented in this study.

In Figure 4, a comprehensive confusion matrix is depicted, assigning specific numerical labels to various text categories, namely: Public Transport Infrastructure (labeled as 0), Socioeconomic Disparity (labeled as 1), Accessibility (labeled as 2), and Others (labeled as 3). This categorization allows for a clear and organized representation of the data, facilitating a deeper understanding of the relationships and patterns within the mentioned text categories. Figure 5 shows the text classification outcomes both in percentage and the actual value in the pie chart.

**Table 2: Sample tweets in each category**

| Tweet Category | Sample Raw Tweets |
|---|---|
| Public Transport Infrastructure | @NYCTSubway L train looks very clean and passenger self-distancing |
| Socioeconomic Disparity | With so many major challenges in our world today, it may seem that tackling transnational #OrganizedCrime remains crucial. However, let's not forget the systemic issue of #SocioeconomicDisparity. It's time we prioritize battling #Inequality alongside #ClimateChange, #Pandemics, and #Migration. Our collective future depends on it. |
| Accessibility | We're committed to making our services more accessible for everyone. With improved accommodation and better access points, we ensure no one is left behind. #AccessibilityMatters |

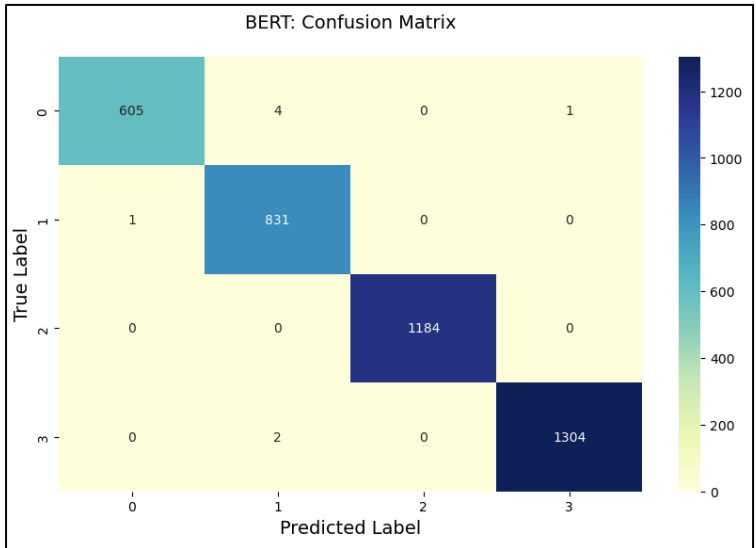
*Figure 4: BERT model performance*

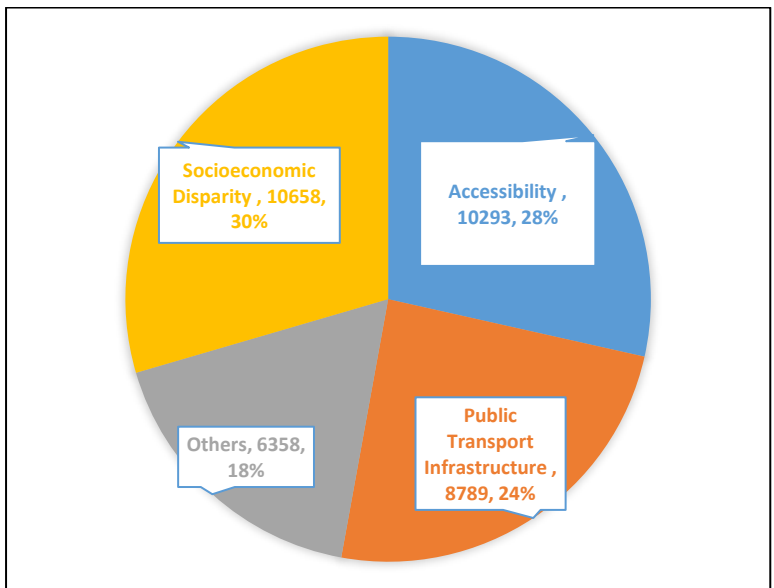
*Figure 5: Text classification outcome*

**Sentiment analysis of public perception**
The heatmap in

Figure 6 provides an insightful visualization of the distribution of tweets concerning various transportation-related topics and their associated sentiments. One of the most prominent observations is the prevalence of certain topics, specially Socioeconomic Disparity and Accessibility. These topics have a higher volume of tweets compared to others, indicating people expressed their concern about these topics.

Further inspection of the sentiment distribution reveals that negative sentiment is prevalent across all topics. This dominance suggests a general trend of dissatisfaction or criticism in tweets associated with these transportation issues. In particular, the topics of *Accessibility* and *Socioeconomic Disparity* not only attract a high volume of tweets, but a significant portion of these tweets also express negative sentiment. This pattern could be indicative of public dissatisfaction with these aspects of transportation. On the other end of the spectrum, topics such as *Public Transport Infrastructure*, and *Others* see relatively fewer tweets. This could imply that these topics are either less controversial, less relevant to the user base, or simply less discussed on Twitter.

When comparing neutral and positive sentiments, it's noticeable that tweets with a positive sentiment are less common. This observation suggest that users are more likely to tweet about these topics when they harbor negative experiences or are indifferent, rather than when they have positive experiences. People generally expressed neutral sentiments when tweeting about *Public Transport Infrastructure*; in other words, the lack of extreme positive or negative sentiment might be a sign that people are generally satisfied with the existing public transportation infrastructure in NYC. They have no significant praises (positive sentiment) or complaints (negative sentiment), hence the majority of the sentiment being neutral. These insights can prove valuable to policymakers, urban planners, and transit agencies, providing a snapshot of public sentiment and areas of concern related to various aspects of transportation.

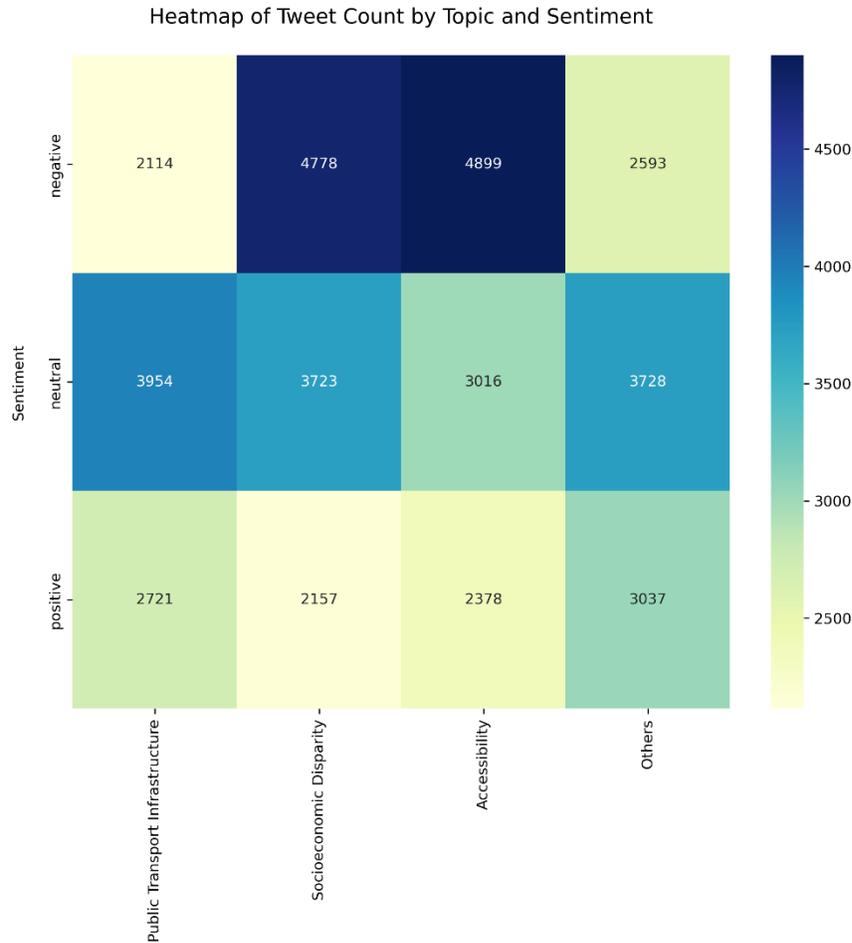

Figure 6: Heatmap of Tweet Count by Topic and Sentiment

**Choice Model Results**

The *Table 3* comprises descriptive statistics for five different metrics associated with transportation, socioeconomic factors, and sentiment analysis. The first metric, Public Transport Infrastructure, has an average value of 0.23 and a standard deviation of 0.421. Socioeconomic Disparity, the second metric, has similar characteristics with a mean of 0.24, a standard deviation of 0.427, and a range from 0 to 1. The third metric, Transport Accessibility, is characterized by a mean of 0.027 and a standard deviation of 0.162, with values ranging from 0 to 1. The considerably low mean, relative to the maximum possible value, suggest that in most of the locations analyzed, transportation is not readily accessible. The final two metrics pertain to sentiment analysis. Positive Sentiment has a mean of 0.367 and a standard deviation of 0.482, while Neutral Sentiment has a mean of 0.365 and a standard deviation of 0.481. Both metrics range from 0 to 1. These represent the proportions of positive and neutral sentiments or reactions towards the transportation system, respectively.

Table 3 Descriptive Statistics of Selected Variables

|  | Mean | Standard Deviation | Minimum | Maximum |
|---|---|---|---|---|
| Public Transport Infrastructure | 0.230 | 0.421 | 0 | 1 |
| Socioeconomic Disparity | 0.240 | 0.427 | 0 | 1 |
| Transport Accessibility | 0.027 | 0.162 | 0 | 1 |
| Positive Sentiment | 0.367 | 0.482 | 0 | 1 |
| Neutral Sentiment | 0.365 | 0.481 | 0 | 1 |
| Negative Sentiment | 0.268 | 0.443 | 0 | 1 |
| Percent unemployed | 2.087 | 1.146 | 0.386 | 14.019 |
| Median income | 126953.415 | 54700.639 | 12073 | 250001 |
| Identified as disadvantaged | 0.049 | 0.215 | 0 | 1 |
| Expected agricultural loss rate (Natural Hazards Risk Index) (percentile) | 10.439 | 13.162 | 0 | 94 |
| Expected building loss rate (Natural Hazards Risk Index) (percentile) | 17.327 | 21.237 | 0 | 99 |
| Energy burden (percentile) | 29.182 | 28.905 | 0 | 96 |
| PM2.5 in the air (percentile) | 49.802 | 6.998 | 31 | 65 |
| Diesel particulate matter exposure (percentile) | 96.848 | 3.881 | 0 | 99 |
| Traffic proximity and volume (percentile) | 78.355 | 8.690 | 0 | 99 |
| Is low income and high percent of residents that are not higher ed students? | 0.045 | 0.207 | 0 | 1 |

**Categorical Variable**

| Gender | Count | Percentage |
|---|---|---|
| Female | 13061 | 36.182 |
| Other | 23037 | 63.818 |

| Race | Count | Percentage |
|---|---|---|
| White | 31431 | 87.071 |
| Asian | 2780 | 7.701 |
| Hispanic | 1256 | 3.479 |
| Black | 584 | 1.618 |
| American Indian | 47 | 0.130 |

**Model Development and Estimation Results**

The table shows three separate discrete choice models. Each model reveals factors affecting individuals' likelihood of tweeting related to transport accessibility, socioeconomic disparity, and public transport infrastructure issues.

**Transport Accessibility**

A binary logit model is estimated to understand the factors affecting individuals' tweets about transport accessibility. There are two choices in this case: the tweet is related to transport accessibility, and the tweet is not related to transport accessibility. A total of 36098 tweets have been included in the final model. The adjusted rho-squared against the null model is 0.81, which is an excellent fit against the null model.

It is observed that females are more likely to discuss transport accessibility issues than males. This study also tested how individuals from different races affect accessibility reporting on Twitter. The model result depicts that individuals of Asian origin are more likely to tweet regarding accessibility than whites. It is more likely to receive negative tweets than neutral ones when discussing transport accessibility.

It is worth mentioning individuals who live below the poverty line, who are identified as disadvantaged, or who are currently unemployed may need more internet access to tweet frequently. The model results support this hypothesis. It is also found that unemployed individuals are less likely to tweet about transport accessibility. It is also found that individuals who are identified as disadvantaged are less likely to tweet about transport accessibility. A similar trend has been found with people living in low-income neighborhoods.

The model result shows that individuals with a total household income of more than $50,000 are more likely to tweet about transport accessibility. It is found that individuals who live in a location where there is a higher volume of traffic are more likely to tweet about transport accessibility. People who live in places with a higher natural hazard risk index are less likely to tweet about transport accessibility. Similar results have been found where individuals live in homes with higher energy burden areas.

**Socioeconomic Disparity**

Two choices have been modeled to understand the factors affecting the tweets related to socioeconomic disparity. The choices include the tweet being related to socioeconomic disparity and the tweet not associated with socioeconomic disparity. The goodness-of-fit against the null model is 0.812, which is a good fit. It is found that females are more likely to tweet about socioeconomic disparity. The model results demonstrate that white individuals are less likely to tweet about socioeconomic disparity. In contrast, it is found that individuals of Asian origin are more likely to tweet about socioeconomic disparity.

The model results reveal that negative tweets are more likely to be related to socioeconomic disparity. It is found that unemployed individuals are less likely to tweet about socioeconomic disparity. Similar results have been found for individuals who are identified as disadvantaged.

The model uses several natural hazard indices, such as expected agricultural loss, building loss, and energy burden. The model results show that those living in those areas are less likely to tweet about socioeconomic disparity. The model shows that individuals experiencing different types of air pollution (i.e., PM2.5, particulate matter, etc.) are more likely to tweet about socioeconomic disparity.

**Public Transport Infrastructure**

A binary logit model has been estimated to model two choices: the tweet is related to public transport infrastructure, and the tweet is not related to public transport infrastructure. The adjusted rho-squared value against the null model is 0.823. The model results demonstrate that unemployed individuals are more likely to tweet about the public transport infrastructure. It is also found that individuals whose median income is more than $50,000 are more likely to tweet about public transport infrastructure.

The model results show that individuals who are identified as disadvantaged are less likely to tweet about public transport infrastructure. It is also found that individuals who live in areas with high air pollution are more likely to tweet about public transport infrastructure.

*Table 4: Binary logit model*

|  | Transport Accessibility | | Socioeconomic Disparity | | Public Transport Infrastructure | |
| --- | --- | --- | --- | --- | --- | --- |
| Number of observations | 36098 | | 36098 | | 36098 | |
| Log-likelihood value of full model | -4659.241 | | -4695.159 | | -4418.215 | |
| Log-likelihood value of null model | -25021.22 | | -25021.220 | | -25021.220 | |
| Adjusted Rho-squared value against null model | 0.813 | | 0.812 | | 0.823 | |
| **Variable** | **Parameter** | **t-stat** | **Parameter** | **t-stat** | **Parameter** | **t-stat** |
| Constant | -3.962 | -36.330 | -3.364 | -40.870 | -3.640 | -52.676 |
| Female (1=Female, 0= Other) | 0.691 | 5.949 | 0.395 | 2.462 | 1.351 | 11.109 |
| Race: White (1=White, 0=Other) | 0.133 | 1.434 | -1.039 | -9.635 | -1.587 | -14.978 |
| Race: Asian (1=White, 0=Other) | 0.523 | 3.565 | 0.065 | 0.424 | -0.329 | -2.173 |
| Sentiment: Neutral | 0.395 | 2.947 | 0.600 | 4.245 | 0.277 | 1.587 |
| Sentiment: Negative | 0.833 | 7.122 | 1.212 | 10.415 | --- | --- |
| Percent unemployed greater than 0.01(%) | -0.295 | -3.248 | -0.584 | -5.628 | 0.319 | 4.466 |
| Median Income more than 50,000 ($) (1=Yes, 0=No) | 0.333 | 2.466 | 1.457 | 11.204 | 1.476 | 11.324 |
| Identified as disadvantages (1=Yes, 0=No) | -0.217 | -0.951 | -0.469 | -1.574 | -0.256 | -1.240 |
| Traffic proximity and volume (percentile) | 0.829 | 9.666 | 0.158 | 1.907 | --- | --- |
| Expected agricultural loss rate greater than 0.01 (Natural Hazards Risk Index) (1=Yes, 0=No) | -0.257 | -2.514 | -0.297 | -3.052 | -0.596 | -7.964 |
| Expected building loss rate greater than 0.01 (Natural Hazards Risk Index) (1=Yes, 0=No) | -0.857 | -8.319 | -0.470 | -5.689 | --- | --- |
| Energy burden greater than 0.01 (percentile) (1=Yes, 0=No) | -0.099 | -1.294 | -0.088 | -1.068 | --- | --- |
| PM2.5 in the air greater than 0.01 (1=Yes, 0=No) | -0.101 | -1.140 | 0.354 | 4.577 | 0.784 | 10.591 |
| Diesel particulate matter exposure greater than 0.01 (1=Yes, 0=No) | 0.925 | 11.526 | 0.863 | 10.875 | 0.391 | 5.863 |
| Is low income and high percentage of residents that are not higher ed students? (1=Yes, 0=No) | -0.316 | -1.238 | --- | --- | --- | --- |

**CONCLUSIONS**

This study presents an innovative approach to how social network platforms can be used to develop mathematical models and evidence-based policies. The study converted texts to survey-type data using natural language processing techniques. In addition, a series of algorithms have been developed to retrieve socioeconomic data from various sources such as Census, Social Security Administration, and climate and economic justice data. After compiling the dataset, a series of advanced discrete choice models are estimated to understand the factors affecting public attitude towards transport accessibility, socioeconomic disparity, and public transportation.

Key findings from this study are outlined as follows.

**Transport Accessibility:**
- *Females are more likely to discuss transport accessibility issues than males.*
- *Individuals of Asian origin are more likely to tweet regarding accessibility than Whites.*
- *Unemployed individuals, individuals who are identified as disadvantaged, and individuals living in low-income neighborhoods are less likely to tweet about transport accessibility.*
- *People are more likely to express negative sentiments while tweeting about transport accessibility.*
- *Individuals who experience high traffic in their neighborhood are more likely to express concern about transport accessibility.*
- *People living in communities with higher natural hazard risk indices are less likely to tweet about transport accessibility.*

**Socioeconomic Disparity:**
- *White individuals are less likely to tweet about socioeconomic disparity as opposed to individuals with Asian origin.*
- *Individuals are more likely to express negative sentiments while tweeting about socioeconomic disparity.*
- *Individuals experiencing different types of air pollution (i.e., PM2.5, particulate matter, etc.) are more likely to tweet about socioeconomic disparity.*
- *Unemployed individuals and individuals who are identified as disadvantaged are less likely to tweet about socioeconomic disparity.*
- *People living in communities with higher natural hazard risk indices (such as expected agricultural loss, building loss, and energy burden) are less likely to tweet about socioeconomic disparity.*

**Public Transportation Systems:**
- *Unemployed individuals are more likely to tweet about the public transportation systems.*
- *Individuals whose median income is more than $50,000 are more likely to tweet about public transportation systems.*
- *Individuals who are identified as disadvantaged are less likely to tweet about public transportation systems.*
- *Individuals who are exposed to high air pollution are more likely to tweet about public transportation systems.*

This study used sentiment analysis in decoding public perceptions on key transportation issues; the high volume of tweets and prevalence of negative sentiment regarding Socioeconomic Disparity and Accessibility underline public dissatisfaction with these aspects, calling for urgent attention from policymakers. Meanwhile, the neutral sentiment dominating Public Transport Infrastructure-related tweets may indicate public satisfaction with the existing framework, a potentially less controversial topic for further discourse. These findings can shape a more responsive and informed approach in transport planning and policymaking, enabling a better alignment with public sentiment and needs. However, the current study explicitly looked at various equity features. It would be interesting to see whether individuals' travel modes, destination locations, departure times, and activity types can be generated using Twitter data. Also, a natural extension would be looking at trip-chain rather than single trips. While this study offers valuable insights, its scope was limited to one week of Twitter data from New York City, which may affect the broad applicability of the findings. A longer data collection period, as well as inclusion of other cities, could offer a more comprehensive understanding of accessibility, public transport infrastructure, and socioeconomic disparity.

**ACKNOWLEDGMENTS**
The material presented in this paper is based on work supported by the National Science Foundation under Grants No. OIA-1946093 and SCC-PG- 2229439. Any opinions, findings, and conclusions or recommendations expressed in this paper are those of the authors and do not necessarily reflect the views of the National Science Foundation.

**AUTHOR CONTRIBUTIONS**
Khondhaker Al Momin: Data collection, Methodology, Formal analysis, Draft preparation, and Review & editing. Arif Mohaimin Sadri: Conceptualization, Draft preparation, Methodology, Supervision, and Review & editing. Md Sami Hasnine: Formal analysis, Draft preparation, and Review & editing. All authors reviewed the results and approved the final version of the manuscript.